\documentclass[modern]{aastex63}
\usepackage{natbib}
\usepackage{rotating}
\usepackage{hyperref}

\begin{document}

\title{A Novel Survey for Young Substellar Objects with the W-band filter  II. The Coolest and Lowest Mass Members of the Serpens-South Star-forming Region}

\correspondingauthor{Jessy Jose}
\email{jessyvjose1@gmail.com}

\author{Jessy Jose}
\affiliation{Indian Institute of Science Education and Research (IISER) Tirupati, Rami Reddy Nagar, Karakambadi Road, Mangalam (P.O.), Tirupati 517 507, India}
\affiliation{Kavli Institute for Astronomy and Astrophysics, Peking University, Yi He Yuan Lu 5, Haidian Qu, Beijing 100871, China}

\author{Beth A. Biller}
\affiliation{SUPA, Institute for Astronomy, University of Edinburgh, Blackford Hill, Edinburgh EH9 3HJ, UK}
\affiliation{Centre for Exoplanet Science, University of Edinburgh, Edinburgh, UK}

\author{Lo\"{\i}c Albert}
\affiliation{Institut de recherche sur les exoplan\`etes, Universit\'e de Montr\'eal, Montr\'eal, H3C 3J7, Canada}

\author{Sophie Dubber}
\affiliation{SUPA, Institute for Astronomy, University of Edinburgh, Blackford Hill, Edinburgh EH9 3HJ, UK}
\affiliation{Centre for Exoplanet Science, University of Edinburgh, Edinburgh, UK}

\author{Katelyn Allers}
\affiliation{Bucknell University; Department of Physics and Astronomy; Lewisburg, PA 17837, USA}

\author{Gregory J. Herczeg}
\affiliation{Kavli Institute for Astronomy and Astrophysics, Peking University, Yi He Yuan Lu 5, Haidian Qu, Beijing 100871, China}

\author{Michael C. Liu}
\affiliation{Institute for Astronomy, University of Hawaii at Manoa, 2680 Woodlawn Drive, Honolulu, HI, 96822, USA}

\author{Samuel Pearson}
\affiliation{SUPA, School of Physics \& Astronomy, University of St Andrews, North Haugh, St Andrews, KY16 9SS, United Kingdom}

\author{Bhavana Lalchand}
\affiliation{Graduate Institute of Astronomy, National Central University, 300 Jhongda Road, Zhongli, Taoyuan 32001, Taiwan}

\author{Wen-Ping Chen}
\affiliation{Graduate Institute of Astronomy, National Central University, 300 Jhongda Road, Zhongli, Taoyuan 32001, Taiwan}

\author{Micka\"el Bonnefoy}
\affiliation{Institut de Plan\'etologie et d'Astrophysique de Grenoble, Universit\'e Grenoble Alpes, CS 40700, 38058 Grenoble C\'edex 9, France}

\author{Etienne Artigau}
\affiliation{Institut de recherche sur les exoplan\`etes, Universit\'e de Montr\'eal, Montr\'eal, H3C 3J7, Canada}

\author{Philippe Delorme}
\affiliation{Institut de Plan\'etologie et d'Astrophysique de Grenoble, Universit\'e Grenoble Alpes, CS 40700, 38058 Grenoble C\'edex 9, France}

\author{Po-shih Chiang}
\affiliation{Graduate Institute of Astronomy, National Central University, 300 Jhongda Road, Zhongli, Taoyuan 32001, Taiwan}

\author{Zhoujian Zhang}
\affiliation{Institute for Astronomy, University of Hawaii at Manoa, 2680 Woodlawn Drive, Honolulu, HI, 96822, USA}

\author{Yumiko Oasa}
\affiliation{Faculty of Education / Graduate School of Science and Engineering, Saitama University, 255 Shimo-Okubo, Sakura, Saitama, Saitama 338-8570, Japan}

\begin{abstract}
Given its relative  proximity ($\sim$430 pc), compact size ($<$ 20$^\prime$), young age ($\sim$ 0.5 Myr) and rich number of 
young stellar objects,   the Serpens-South star forming region is a promising site for studying young sub-stellar objects, 
yet the low-mass members of this region remain largely undiscovered.  
In this paper we report on a deep photometric survey  using a custom 1.45 $\mu$m filter ($W$-band),  as well as standard $J$ and $H$ near-IR filters, in order to identify candidate low-mass young brown-dwarfs in the Serpens-South region.  We constructed a reddening-insensitive index ($Q$) by combining $J$, $H$ and $W$-band photometry for survey objects, in order to identify candidate low-mass members of Serpens based on the strength of the water absorption feature at 1.45 $\mu$m in the atmospheres of mid-M and later objects.  We then conducted spectroscopic follow up to confirm youth and spectral type for our candidates.  This is the first survey to identify the very low-mass and coolest members of Serpens-South. We identify 4 low-mass candidate Serpens members, which all display IR excess emission, indicating the likely presence of circumstellar disks around them. One of the four candidate low-mass members in our list, SERP182918-020245, exhibits $Pa{\beta}$ and $Br{\gamma}$  emission features, confirming its youth and ongoing magnetospheric accretion. Our new candidate members have spectral types $>$M4 and are the coolest and lowest mass candidate members yet identified in Serpens-South. 
\end{abstract}


\section{Introduction}
\label{intro}

Around a dozen young, giant exoplanets have been imaged to date, with estimated masses of 5-13 M$_{Jup}$ and ages less than $<$100 Myr \citep{bowler2016}. 
These objects require extremely high contrasts to image and are difficult to characterize, given their proximity 
to bright host stars.    Since substellar objects  are 
significantly brighter and warmer at very young ages, i.e., less than 10 Myr \citep{chabrier2000}, the detection of  free-floating sources with a few 
Jupiter masses is possible by exploring nearby star-forming regions (\citealt{oasa1999,lucas2000}). These objects are key analogues to directly imaged exoplanets and 
can be studied in detail, as they are not obscured by a nearby bright star.  They offer the chance to study planetary mass objects in 
the moments of their formation, critically constraining possible formation mechanisms.  Detecting and characterizing substellar and 
planetary mass objects in these regions not only helps to characterize the atmospheric properties of these objects, but tests substellar evolutionary models at the youngest ages.    

Detecting a statistically-significant population of substellar and planetary mass objects in young star-forming regions also critically 
constrains the initial mass function (IMF), the  distribution of stellar mass in a star forming event.  The IMF  is mainly determined 
by the star formation process itself. Constructing and characterizing the  complete sample of substellar members  in various star 
forming regions is  fundamental to understand the role of environment in their formation and evolutionary processes (i.e., minimum mass, 
disk evolution etc). Hence by estimating the form of the IMF of a given event,  theories of the star formation can be tested.  Various 
theories predict a large range for the minimum masses of the IMF ($\sim$ 1-100 M$_{Jup}$, \citealt{low1976,larson1992,whitworth2007}, 
references therein) and this topic has been highly debated over the last several decades. 

There have been numerous attempts in the past to detect and characterize the substellar and planetary-mass object in nearby star 
forming regions at distances of  $\sim$ 100--400 pc. These searches are mainly based on  optical-infrared photometry and astrometry 
using  wide-field surveys such as PanSTARRS1, SDSS, 2MASS, UKIDSS, {\it Spitzer Space Telescope}, WISE and GAIA (\citealt{cutri2003,lucas2008}). Follow-up spectroscopic observations in optical-IR are used to constrain the physical parameters of the  candidate low-mass objects.  Taurus, 
Perseus,  IC 348, NGC 1333, and Upper-Scorpius are some of the nearby star forming regions which have been extensively surveyed by many 
authors down to a few Jupiter masses (e.g. \citealt{kraus2017,esplin2017,best2017,lodieu2017,zhang2018}, many more). %

The Serpens Molecular Cloud is one of the most active sites of ongoing star formation within 500 pc (see \citealt{herczeg2019,eiroa2008} and references therein).  The Serpens-South protocluster is located at 
 the center of a well-defined filamentary cloud and is  a part of the  larger Serpens-Aquila rift molecular cloud complex  (\citealt{bontemps2010,gutermuth2008,andre2010}). Given its relative  proximity
 ($436$ $\pm$ $9.2$ pc, \citealt{ortiz2017}, see also analysis of Gaia DR2 by \citealt{ortiz2018} and \citealt{herczeg2019}), 
 compact size ($<$ 20$^\prime$), young age ($\sim$ 0.5 Myr) and rich  population of young stellar objects (protostellar fraction $\sim$ 80-90\%, \citet{gutermuth2008}),  the Serpens-South star  forming region is a   promising site for studying young sub-stellar objects.    Since its discovery, Serpens-South has become the center of a wide range of 
 observational analysis including  near, mid, far-infrared and X-ray mappings with {\it Spitzer},{\it Herschel} and {\it Chandra} 
 that trace  heated  dust around protostars (\citealt{gutermuth2008,bontemps2010,povich2013,dunham2015,getman2017}), millimeter mappings   revealing cold dust \citep{maury2011}, 
 near-infrared polarimetry revealing the importance of global magnetic fields in the cluster formation history \citep{sugitani2011},   molecular outflows studies (e.g., \citealt{nakamura2011,teixeira2012,plunkett2015a,plunkett2015b}), 
 and a wealth of spectral line 
 surveys probing filamentary infall (e.g., \citealt{friesen2013,kirk2013,tanaka2013,fernandez2014,nakamura2014}).

In spite of the wealth of optical, infrared and submillimeter surveys currently available 
towards Serpens-South, there have been no dedicated searches reported in the literature towards its very low-mass members.
 The latest search for low-mass  members of the region by \cite{winston2018} using {\it Chandra}, 2MASS and {\it Spitzer} photometry identifies 66 young stellar objects for which the detection limit  hardly reaches   down to $\sim$ 0.1 M$_{\odot}$ (see Fig. \ref{jhj_cmd}), implying that a large fraction of  sub-stellar objects are still   undetected  in  this region.  
 
 Canonical approaches to sub-stellar studies rely on photometric identification followed by spectroscopic  confirmation.  
 However,  because of the non-uniform  reddening of the medium ($A_V$ $\sim$ 5-75 mag, \citet{povich2013}), the low-mass members are difficult to distinguish from  reddened background stars using traditional photometric methods.    Also, optical/NIR spectroscopy is time consuming  for very faint, highly extincted  star forming regions such as Serpens-South.  An efficient manner of searching for these  young low-mass objects is to identify them via unique spectral features, in particular  the 1.45 $\mu$m H$_2$O absorption band seen in the spectra of M-L-T-Y type objects. Photometry with a narrow band  filter centered on
the deep H$_2$O absorption feature seen at  1.45 $\mu$m  in the spectra of young brown dwarfs (Late M and L type spectral types) can break
the degeneracy in colors between the young brown dwarfs and reddened background stars, which have no  detectable water absorption. 
We can combine 1.45 $\mu$m imaging with $J$ and $H$ photometry to robustly estimate spectral types independent of reddening 
(Allers \& Liu, submitted). 

In this paper we report on a deep photometric survey  using the 1.45 $\mu$m ($W$-band),  $J$ and $H$-bands  in order to identify 
candidate young brown dwarfs and very low mass stars in the Serpens-South region, as well as spectroscopic follow up and confirmation of these 
candidates. This is the first survey to identify the very low-mass and coolest members of Serpens-South. Of the $\sim$ 69 candidate low-mass 
objects selected based on 1.45 $\mu$m, $J$ and $H$-band photometry, we obtained spectra for 7 targets.  Four targets are confirmed 
as spectral  type $\ge$ M4 objects hosting circumstellar disks.  These objects are indeed the coolest and lowest mass members yet detected in Serpens-South star forming region.

\section{OBSERVATIONS AND DATA ANALYSIS}

\subsection{CFHT-WIRCam $J$, $H$ and $W$-band data sets}
\label{WIRCam}

\begin{figure*}
\centering
\includegraphics[scale = 0.5, trim = 50 270 70 20, clip]{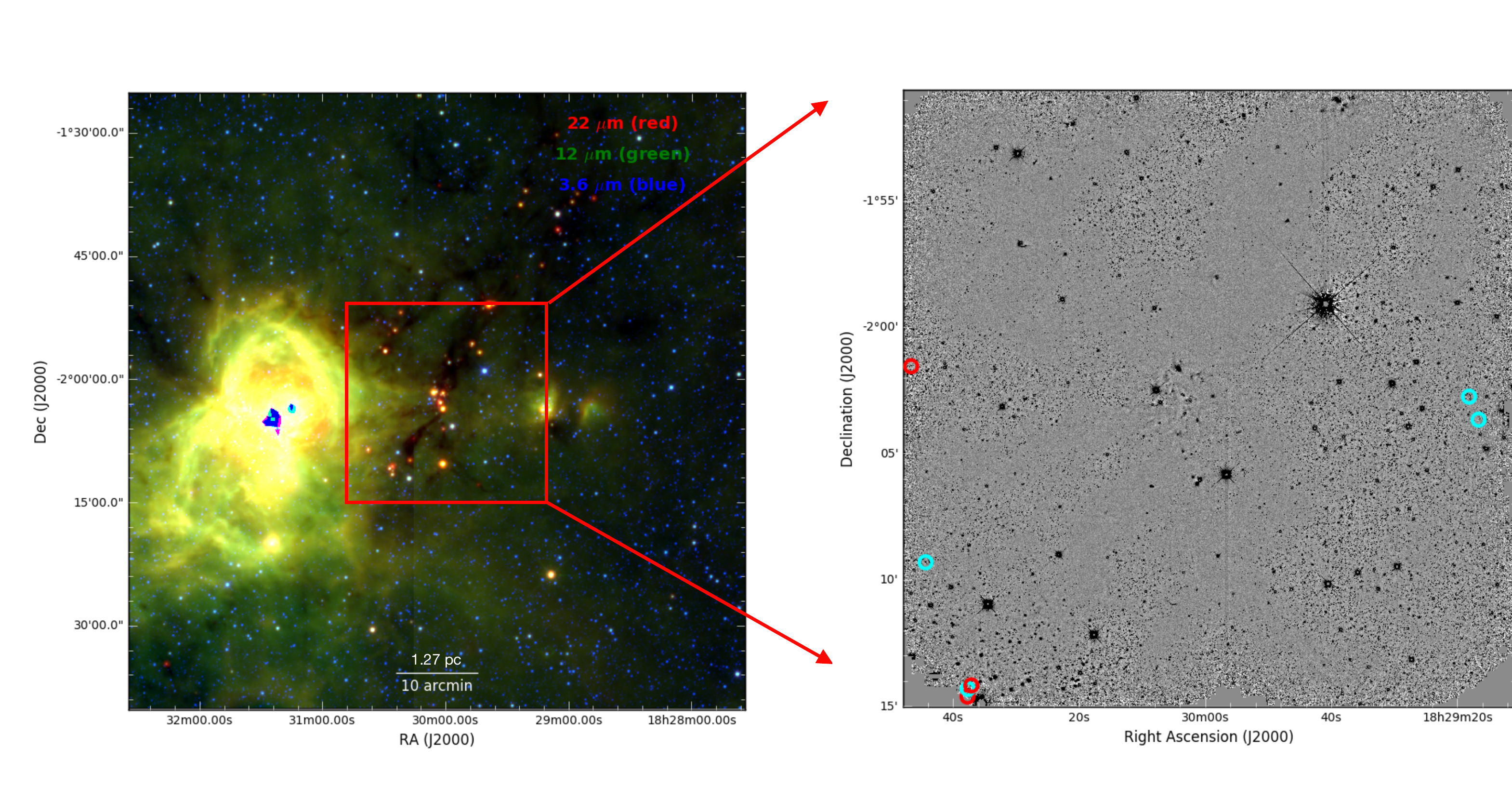}
\caption{{\it left:} Color-composite image of Serpens-South and W40 region obtained using images at 22 $\mu$m (red), 
12 $\mu$m (green) and 3.6 $\mu$m (blue)   from WISE archive.   The red box indicates the area covered by the WIRCam observations, 
which essentially covers the dark filament through  Serpens-South.  {\it right: } $H$-band image of   WIRCam  and the 
candidate low-mass objects selected for spectroscopic follow-up are shown in circles, where, the confirmed low-mass members are shown in cyan (see Section \ref{results} for details).  }
\label{fov}
\end{figure*}
 Over the last 5 years, our team has conducted a  large observing program to detect and characterize the low-mass young stellar
 population of nearby star forming regions using the Wide Field IR Camera \citep[WIRCam,][]{puget2004} at the Canada France Hawaii 
 Telescope (CFHT) with a custom-built medium-width (6\%) filter centered on a water-absorption feature at 1.45$\mu$m (henceforth 
 the $W$-band filter).  In this paper we report the results of observations on 2016 July 14-15 of the Serpens-South region using 
 the $W$-band filter  along with  the broad band  WIRCam $J$ and  $H$ filters.  WIRCam has a field-of-view  of $\sim$ 20$^\prime$ 
 $\times$ 20$^\prime$ with a sampling  of 0.3 arcsecond per pixel. A single WIRCam pointing  centered at RA = 18$^h$30$^m$03$^s$ 
 and Dec = -02$^d$01$^m$58$^s$ was sufficient to cover the Serpens-South cluster. We used a 21-point dithering 
 pattern to fill the gaps between the four detectors of WIRCam and to accurately subtract the sky background. The total integration 
 time for $J$, $H$ and $W$-band filters were 1890, 1920, and 12285s, respectively. Pre-processing of the images were performed for 
 dark frame subtraction, flat fielding, sky subtraction, bad pixel masking and astrometric calibration. After combining the  
 individual exposures into final stacked images in each filter, the source extraction and photometry were performed by using 
 SExtractor \citep{bertin1996}. In order to avoid  artifacts and false detections, we selected only those sources having
S/N $>$ 10. Absolute photometric calibration in $J$ and $H$ bands  was performed using data from the 2MASS point source catalog 
\citep{skrutskie2006}. To calibrate our $W$-band photometry, we first create a catalog of available photometry for objects in 
our field from the $WISE$, $SDSS$, $DENIS$, $2MASS$, $UKIDSS$, $APASS$, and $USNO-B1$ catalogs.  For any object in our field having available near-IR and additional (shorter or longer wavelength) broad-band photometry, we 
determine a best-fit, reddened template spectrum from the SpeX Spectral Library \citep{rayner2009, cushing2005}.  We first scale 
the synthetic and observed photometry of the templates to match the observed $J$ and $H$-band magnitudes and then determine the closest matching reddened template to all photometry.  We then adopt a likely $W$-band magnitude for the object in our field as the synthetic, scaled $W$-band 
photometry of the best-fit template.  By comparing each object's likely $W$-band flux (based on template fitting) to the measured 
flux in the CFHT $W$-filter frame, we calculate a likely $W$-filter zero point.  We then use the median of these likely photometric 
zero points to calculate $W$-band magnitudes for all of the objects in the field.  The median absolute deviation of the 1033 
predicted photometric zero points for our field is 0.08~mag, and the standard deviation of the mean is a mere 0.004~mag.  The final catalog after combining the 
photometry from $J$, $W$ and $H$ bands include photometry for 15277 sources, within an uncertainty limit of $<$ 0.1 mag.

\subsection{Selection of candidate low-mass objects}

To distinguish young brown dwarfs and planetary mass objects from reddened background objects, which plague 
surveys based on broad-band colors alone, we followed the approach of Allers \& Liu (submitted) and constructed a reddening-insensitive index ($Q$) by combining $J$, $H$ and $W$-band photometry for survey objects:

\begin{equation}
Q =(J-W) + e \times (H-W)
\end{equation}

where the constant $e$ is expressed in terms of extinction values in each of the three bands:

\begin{equation}
e=(A_J-A_W) / (A_H-A_W).
\end{equation}

\begin{figure}[h]
\centering
\includegraphics[scale = 0.75, trim = 20 0 0 0, clip]{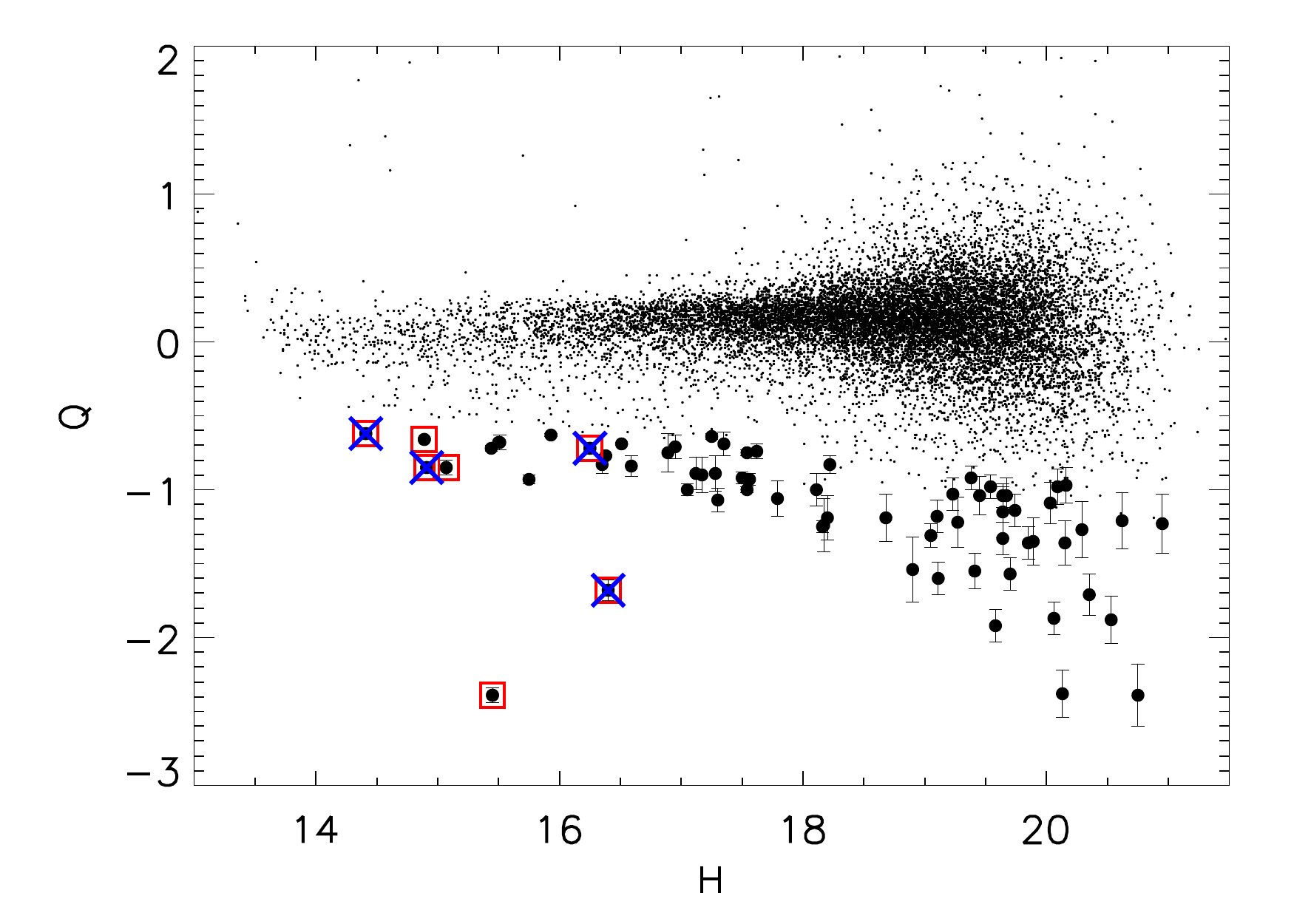}
\caption{$Q$ versus $H$-magnitude distribution of all the sources detected in Serpens-South region.  The sources with  
$H$ $<$ 18 mag, Q  $<$  -( 0.6 +  $\sigma_{Q}$)  and $H>$ 18 mag, Q  $<$  -( 0.6 + 3 $\cdot$ $\sigma_{Q}$)
are highlighted using solid dark circles along with their error bars. Targets for the Spectroscopic observations using ARCoIRIS are marked as  red squares and  
blue crosses represent the confirmed low-mass members of the region (see text for details). 
}
\label{qh}
\end{figure}

Based on synthetic photometry using a M0 spectrum from \citet{kirkpatrick2010}, we adopt e = 1.85, the appropriate value for an M0 star reddened by A$_V$ = 10 for R$_V$ = 3.1 (Allers \& Liu, submitted).  Our typical contaminant is a reddened background M0 star, which should have $Q\sim0$ due to a lack of steam absorption in their spectra.   As expected, the majority of objects detected in our field have a Q-value of $\sim$0 (Figure \ref{qh}).  Later spectral type objects with deepening water absorption features will have progressively more negative values of Q.

Based on synthesized  $J$, $H$ and $W$-band  photometry of young objects (\citealt{allers2013,muench2007}, Allers \& Liu, submitted) and field dwarfs \citep{cushing2005}, low mass stars and brown dwarfs with spectral types later than M6 have $Q$ values $<$-0.6, whereas earlier spectral type contaminants will have $Q$ values $>$-0.6.  The uncertainty on $Q$, $\sigma_Q$ is estimated from propagating the errors from partial derivatives of $Q$ as a function of $J$, $W$ and $H$: 

\begin{equation}
\sigma_Q =\sqrt{\sigma_J^2+(e \cdot \sigma_H)^2 +((1+e) \cdot  \sigma_H)^2}
\end{equation} 

where $\sigma_J$, $\sigma_H$, and $\sigma_W$ are the uncertainties on the $J$, $H$, and $W$-band photometry respectively.

\begin{figure}
\centering
\includegraphics[scale = 0.9, trim = 0 0 200 0, clip]{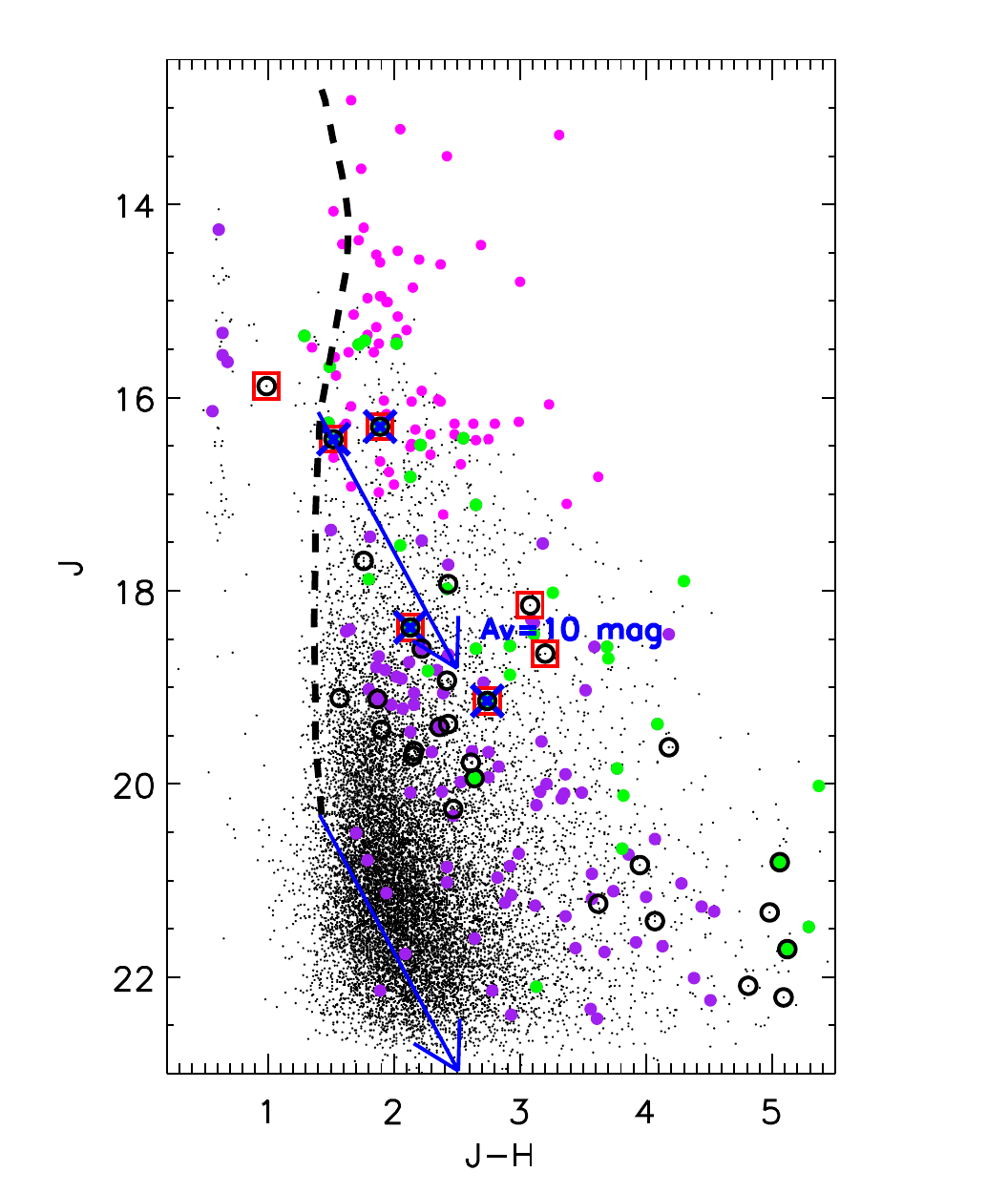} 
\caption{$(J-H)$ vs. $J$   color-magnitude  diagram  for  all the sources (black dots) detected in  CFHT-WIRCam survey of Serpens-South with  photometric uncertainty $<$ 0.1 mag.  The pink sources are from the latest   survey for young stellar objects towards Serpens-South 
using {\it Chandra}, 2MASS and {\it Spitzer} data sets by Winston et al. (2018),   the purple sources are   the probable   members of Serpens-South from the SFiNCs Xray-Infrared catalog by \citet{getman2017} and the green dots   are the young stellar objects in the region from the Gould Belt Survey \citep{dunham2015}.   The black open circles  are those sources which  satisfy the $W$-band   based selection  criteria for 
low-mass objects, i.e. $Q$ $<$  -( 0.6 +  $\sigma_{Q}$) and $H< 18$ mag. Those 7 sources which are  followed up for spectroscopic 
observations are shown as red squares and blue crosses represent the confirmed low-mass members of the region (see text for details).
The dashed curve corresponds   to pre-main sequence isochrone for 2 Myr 
corrected for a distance of 436 pc with solid lines represent the extinction   vectors ($A_V$ = 10 mag) correspond  to 0.1 M$_{\odot}$ 
and 0.01 M$_{\odot}$,   respectively.   
}
\label{jhj_cmd}
\end{figure}

To select high-confidence candidate low mass Serpens-South members, we  considered as candidates those targets with either $H$ $<$ 18 mag, Q  $<$  -( 0.6 +  $\sigma_{Q}$)  or $H>$ 18 mag, Q  $<$  -( 0.6 + 3 $\cdot$ $\sigma_{Q}$).  As photometric errors are significantly larger for fainter objects (as well as the pool of reddened background stars that might act as contaminants), we applied a more stringent cut for objects with $H$ $>$ 18 mag. Thus we obtained 69  candidate low-mass objects  
that satisfy the above criteria. Of these, 31 sources are brighter than 18 mag in $H$-band  and have photometric uncertainties $<$ 0.1 mag 
in all three bands. Fig. \ref{jhj_cmd} shows the $J-H$ vs. $J$ distribution of all the sources detected with CFHT along with the  31 low-mass candidates satisfying  the above selection criteria (i.e, $H$ $<$ 18 mag and  Q  $<$  -( 0.6 +  $\sigma_{Q}$)). Fig. \ref{qh} shows the  $Q$ value distribution of above selected 69 sources with respect to $H$-magnitude.

\subsection{Follow-up spectroscopic Observations}


\begin{table*}
\centering
\caption{Log of spectroscopic observations}
\label{obslog}

\begin{tabular}{ccccc}
 \hline
ID & Name & $\alpha_{(2000)}$& $\delta_{(2000)}$ & Exp. Time(s)   \\
   &      &      deg         & deg               &              \\
\hline
1 & SERP183038-021437 & 277.657049 & -2.243708  & 4 $\times$ 180\\
2 & SERP183038-021419 & 277.656886 & -2.238678 & 8 $\times$ 180\\
3 & SERP183037-021411 & 277.654565 & -2.236360  & 6 $\times$ 120\\
4 & SERP183044-020918 &  277.684766 & -2.155125 & 8 $\times$ 90\\
5 & SERP182917-020340 & 277.319337 & -2.061048 &4$\times$ 180 \\
6 & SERP182918-020245 & 277.325600 & -2.045856  & 12 $\times$ 180\\
7 & SERP183047-020133 & 277.694347  & -2.025817 & 4 $\times$ 90\\

\hline
\end{tabular}
\end{table*}


We performed follow-up spectroscopic observations of 7  bright targets from the list of low-mass candidates using ARCoIRIS 
on the 4 m Blanco telescope at CTIO on the nights of 2016 August 19 and 2016 August 22. The spectra cover a wavelength 
range of 0.8 - 2.47 $\mu$m at a
spectral resolution of R=3500.  We used a standard ABBA nodding sequence along the slit to record object and sky spectra. 
The log of spectroscopic observations is given in Table \ref{obslog}. 
Flat-field and argon lamps were taken immediately after each sets of target.  For each  target we observed a nearby 
A0V star for telluric correction and flux calibration.

\begin{figure*}
\centering
\includegraphics[scale = 0.8, trim = 0 0 0 0, clip]{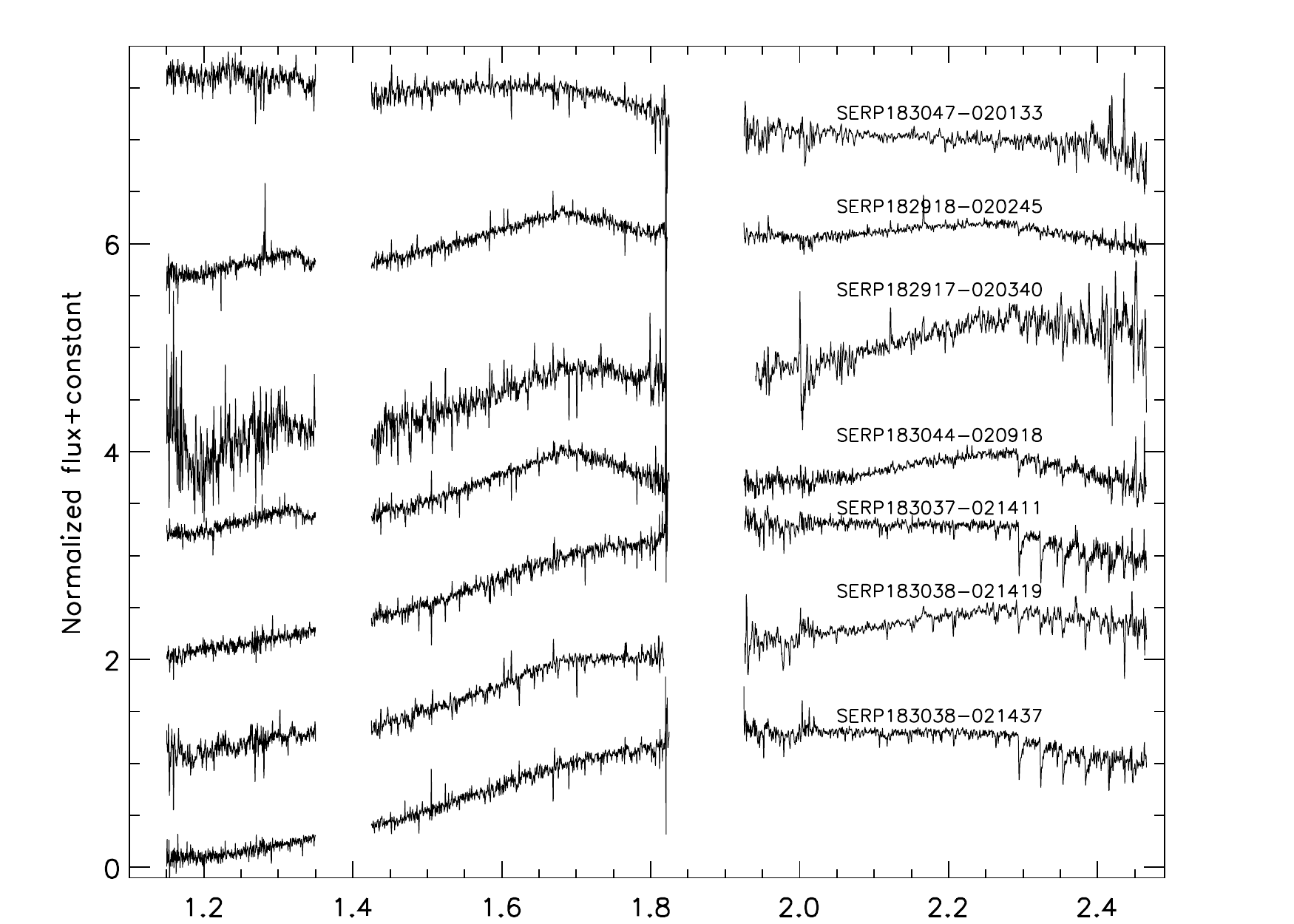}
\caption{Normalized near-infrared spectrum of 7 bright candidate low-mass objects in our list obtained from ARCoIRIS.  }
\label{spec}
\end{figure*}

All reductions of the data taken with ARCoIRIS were carried out with a modified version of the SPEXTOOL package version 4.1 (\citealt{vacca2003,
cushing2004, kirkpatrick2011}).
The individual extracted and wavelength-calibrated spectra from a given sequence of observations, each with their own A0 standard, were then scaled to a 
common median flux and were median combined using $XCOMBSPEC$ in SPEXTOOL. The combined spectra were corrected for telluric 
absorption and were flux-calibrated using the respective telluric standards with $XTELLCOR$. All calibrated sets of observations 
of a given object were then median-combined to produce the final spectrum. Normalized,  smoothed spectra, offset by a constant,  
for the 7 brighter targets in our list is shown in Fig. \ref{spec}.

\subsection{Multi-band data from other catalogs}

\begin{deluxetable}{cccccchhhhhhhhhhhh}
\centerwidetable
\tablecaption{CFHT photometry data for the candidate low-mass objects\label{table:phot}}
\tabletypesize{\normalsize}
\tablehead{
\colhead{ID} & \colhead{Name} & \colhead{[1.45]} & \colhead{$J_{cfht}$} & \colhead{$H_{cfht}$} & \colhead{Q}} 
\startdata
1 & SERP183038-021437&  17.41$\pm$0.01 & 18.65$\pm$0.04 & 15.45$\pm$0.01 & -2.39 $\pm$ 0.05 &  17.017&	14.328&		12.907&	   13.137&    12.180&	11.770&	 9.249&  7.994&	 11.93&  11.80&	  11.59&  11.72\\
2 & SERP183038-021419&  17.954$\pm$0.01 & 19.14$\pm$0.04 & 16.40$\pm$0.01 & -1.68 $\pm$ 0.07 &      18.252&	15.590&		13.910&	   14.201&    12.830& 	11.765&	 9.529&  7.325&	 12.04&  11.31&	  10.79&  10.35 \\
3 & SERP183037-021411&  16.45$\pm$0.01 &  18.15$\pm$0.03 &  15.07$\pm$0.01 & -0.85 $\pm$ 0.05 &      16.934&	14.601&		13.186&	   13.218&    12.580& 	12.271&	 9.939&  7.416&  12.25&  12.10&	  11.91&  12.02 \\
4 &  SERP183044-020918&  15.29$\pm$0.01 &  16.30$\pm$0.01 &  14.41$\pm$0.01 & -0.62 $\pm$ 0.01 &      16.587&	14.304&		13.156&	   13.049&    12.200& 	11.331&	 7.464&  4.237&	 11.94&  11.46&	  11.01&  10.53 \\
5 &  SERP182917-020340&  17.25$\pm$0.01 &  18.38$\pm$0.01 &  16.25$\pm$0.01 &-0.72 $\pm$ 0.02  &      -	&	-     &  	-     &    14.643&    -	    &	-     &   -   &    -  &	 12.87&  11.90&	  11.19&  10.44 \\
6 & SERP182918-020245&  15.74$\pm$0.01 &  16.43$\pm$0.01 &  14.91$\pm$0.01 & -0.85 $\pm$ 0.01 &      16.619&	15.103&		14.167&	   13.943&    12.394& 	11.838&	 8.781&  6.252&	 13.05&  12.36&	  12.23&  11.33 \\
7 & SERP183047-020133&  15.47$\pm$0.01 &  15.88$\pm$0.01 &  14.89$\pm$0.01 & -0.66 $\pm$ 0.01 &      15.845&	14.891&		14.499&	   14.438&    14.094& 	13.419&	 7.882&  5.262&	 14.24&  14.05&	  -    &  -  \\
\enddata
\end{deluxetable}

\begin{deluxetable}{cccccccccccc}
\centerwidetable
\tablewidth{12in}
\tablecaption{Near-infrared and mid-infrared photometry data for the candidate low-mass objects\label{table:litphot}}
\tabletypesize{\footnotesize}
\tablehead{
\colhead{ID} & \colhead{Name} & \colhead{$K_{2mass}$} & \colhead{$K_{ukidss}$} & \colhead{W1} & \colhead{W2} & \colhead{W3} & \colhead{W4} & \colhead{$[3.6]$} & \colhead{$[4.5]$} & \colhead{$[5.8]$} & \colhead{$[8.0]$}} 
\startdata
1 & SERP183038-021437& 	12.907&	   13.137&    12.180&	11.770&	 9.249&  7.994&	 11.93&  11.80&	  11.59&  11.72\\
2 & SERP183038-021419& 	13.910&	   14.201&    12.830& 	11.765&	 9.529&  7.325&	 12.04&  11.31&	  10.79&  10.35 \\
3 & SERP183037-021411&  13.186&	   13.218&    12.580& 	12.271&	 9.939&  7.416&  12.25&  12.10&	  11.91&  12.02 \\
4 &  SERP183044-020918& 13.156&	   13.049&    12.200& 	11.331&	 7.464&  4.237&	 11.94&  11.46&	  11.01&  10.53 \\
5 &  SERP182917-020340& \nodata     &    14.643&   \nodata & \nodata     &   \nodata   &   \nodata  &	 12.87&  11.90&	  11.19&  10.44 \\
6 & SERP182918-020245&  14.167&	   13.943&    12.394& 	11.838&	 8.781&  6.252&	 13.05&  12.36&	  12.23&  11.33 \\
7 & SERP183047-020133&  14.499&	   14.438&    14.094& 	13.419&	 7.882&  5.262&	 14.24&  14.05&	  \nodata    &  \nodata  \\
\enddata
\end{deluxetable}


We identified all matches for the 7 bright targets followed up spectroscopically from the 2MASS \citep{cutri2003}, UKIDSS \citep{lucas2008}, 
{\it Spizer}  and {\it WISE} catalogs. Several of our objects are too faint to have reliable 2MASS $J$ and $H$-band photometry, thus we report only 2MASS $K$ magnitudes here.  Only $K$-band data is available in the UKIDSS catalog. {\it Spitzer}-IRAC four-band data
is obtained from \citet{kuhn2013}. The WISE photometry  at 3.4 (W1), 4.5 (W2), 12 (W3), and 22 $\mu$m (W4) bands are obtained 
from  the All-WISE Source Catalog \citep{cutri2013}. Our CFHT $J$, $H$, and $W$-band photometry for the 7 targets is given in Table~\ref{table:phot} and literature near-infrared and mid-infrared photometry from various catalogs is presented in Table~\ref{table:litphot}.

\section{Results and Discussion}
\label{results}

\subsection{Dereddening and spectral type estimates}
\label{reddening}

\begin{figure}
    \centering
    \includegraphics[scale=0.7, angle=0]{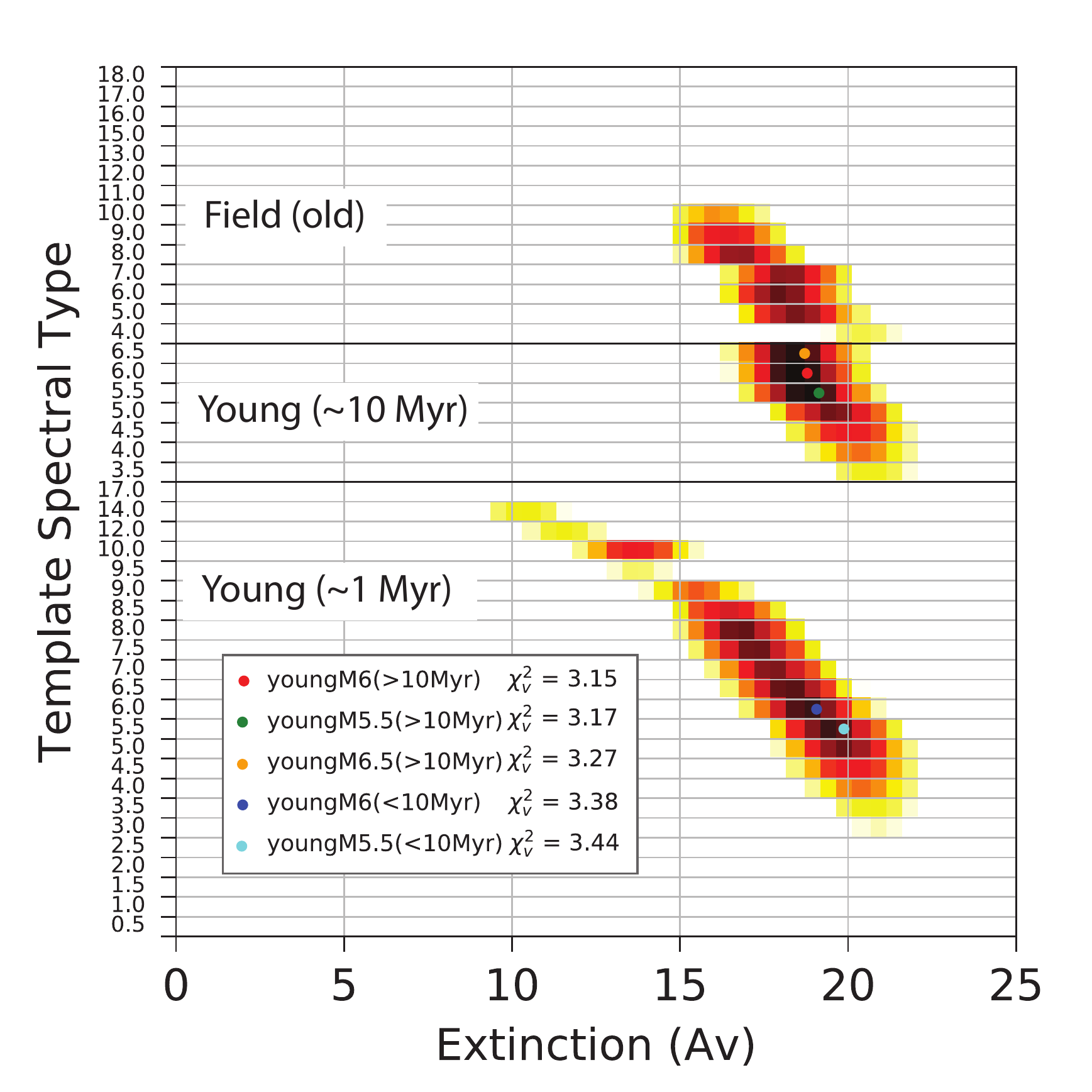}
    \caption{An example of a probability map for SERP182917-020430. These maps were used as tools for determining the adopted spectral type and extinction for each target. We considered 3 distinct sets of standard comparison stars, with ages of $\sim$1 Myr (bottom), $\sim$10 Myr (middle), and old field (top) respectively. 
    Darker regions delineate areas of higher probability. The 5 best-fit spectral type / $A_V$ values are plotted as colored circles.  Numeric spectral types run from M0.5 (0.5) to L6 (16).}
    \label{prob_map}
\end{figure}

We estimate spectral types for our candidates by fitting them with ultra-cool young spectral standards from \citet{luhman2017}, as 
well as old, field spectral standards taken from the SpeX prism library, taking $A_V$, spectral type, and age as free parameters 
(by utilizing separate sets of young vs. old spectral standards).  Our objective is to provide a classification based on the 
pseudo-continuum shape rather than the individual atomic and molecular features.  We calculate reduced $\chi^2$ goodness of fit values for 
a grid of standards at 3 ages: $\sim$1 Myr young spectral standards, somewhat older $\sim$10 Myr spectral standards, and old field spectral standards.  Along this grid, spectral type ranged from M0 to L6 and we reddened templates to $A_V$ values ranging from 0 to 30, using the reddening law from \citet{cardelli1989}. 
These fits are calculated  within the 1.4 - 1.8 $\mu$m  ($H$-band) and 1.90 - 2.30 $\mu$m ($K$-band)
spectral windows, avoiding the noisy 1.8-1.9 $\mu$m range between $H$- and $K$- band.   The J-band window is excluded from the fit 
because the S/N is usually low because of high extinction. 
To obtain a range of best matching templates and corresponding spectral types and extinction, we use $\chi^2$ maps, an example of which is shown in Figure \ref{prob_map}, for SERP183038-021419. To generate these, a reduced $\chi^2$ value is calculated for each grid point, and then converted to a normalised probability (taking the prior that the true spectral type and $A_V$ lie within the ranges for these parameters that we have considered). We used the resulting maps to  identify the areas of parameter space with the highest probability, and used the extent of these areas to estimate errors in spectral type and $A_V$.  In Figure~\ref{prob_map}, regions on the map shown in darker colors delineate the best fitting combinations of spectral type and $A_V$.  For this particular example, the 5 best-fitting models have spectral types of M5.5-M6.5, ages $\leq$10 Myr, and $A_V$ values of 17-19.  However, models with spectral types between M4-M7, ages $\leq$10 Myr, and $A_V$ values of 17-22 provide comparably good fits, with significant degeneracy between spectral type and $A_V$, hence we conservatively adopt this range of spectral types and $A_V$ for this object.

\begin{figure*}
\centering
\includegraphics[scale=0.8, angle = 0]{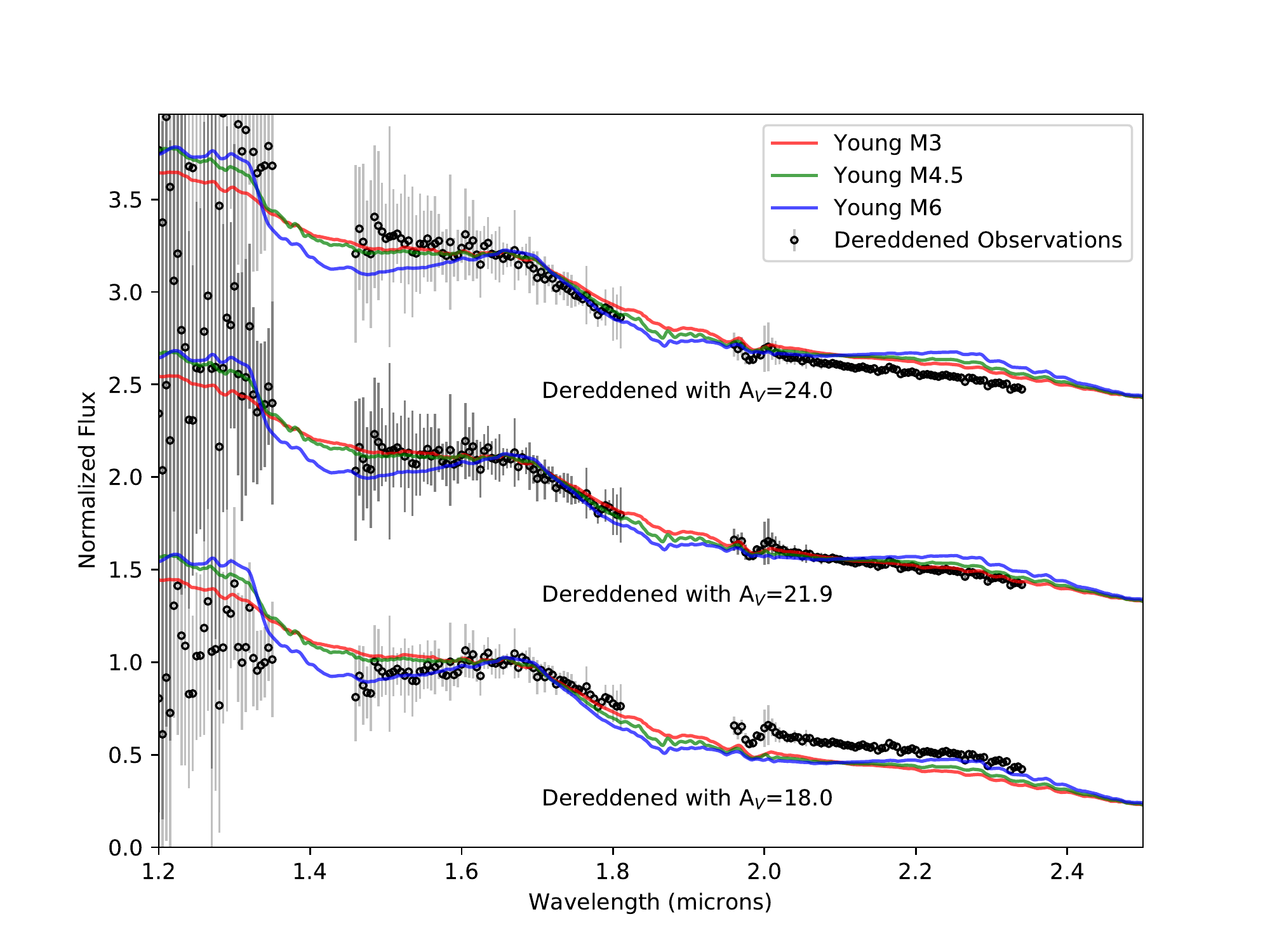}
\caption{ Normalized, dereddened spectra of SERP183038-021419 along with  three best fit  spectral templates and the respective 
$A_V$ values.  Spectra has been binned onto a regular grid sampled at every 5 nm, down from roughly 0.2 nm at full resolving power. The grey error bars show the 1-sigma uncertainty of the binned data. }
\label{spec_2}
\end{figure*}

For late M and L type objects, we can distinguish between old and young objects ($\leq$10 Myr) via the distinctive peaky H-band feature seen in low-surface gravity objects and the shape of the K band continuum \citep{lucas2001, allers2007,rice2010,allers2013,liu2013, schneider2014,gagne2014,esplin2017}, however, our ArcoIRIS spectra do not have the requisite resolution to reveal other spectral indicators of low surface gravity.  No clear spectral indicators distinguishing young and old M0-M4 objects are available in our spectra.  Of the 7 candidate low-mass objects, four are in the spectral range M4-M9, with the latest one being an M7-M9 type, and three  are likely early M  type  ($<$M4) contaminants.  Additionally, the four targets with spectral types $>M4$, SERP183044-020918, SERP182918-020245, SERP183038-021419 and SERP182917-020340 are listed as young stellar objects in the MYStIX IR-Excess Source catalog by \citet{povich2013}, SFiNCs Xray-Infrared catalog by \citet{getman2017} as well as in Gould Belt Survey by \citet{dunham2015}. Evidence of  IR excess  emission from these various catalogs confirm the youth and likely membership of our $>M4$ candidate low mass targets in Serpens-South.  The 
remaining three sources do not appear in any of the above catalogs. 
While we cannot robustly determine age for $<$M4 objects, given the high frequency of field early M dwarfs and lack of other evidence of youth, we do not assign Serpens membership to them.  They are most likely field-age foreground objects.  Figs. \ref{spec_2} to \ref{spec_6} show the normalized, dereddened spectra of four low-mass, young member candidates (see below) of Serpens-South  along with their best matching standard star spectra for three sub-classes and their respective best fit $A_V$ values.  The best fitting spectral types and extinction ranges for our four new candidate Serpens South members are listed in Table \ref{phypars}. Our spectral type classifications have an average uncertainty of $\pm$ 1-3 sub-classes. 

 Given the high extinction of these new cool dwarfs in Serpens-South, we also  implemented spectral fitting using various other extinction laws in the literature such as  \citealt{fitzpatrick1999,fitzpatrick2007}  and also  a steeper  power law with alpha=-2.1 \citep{stead2009}, to determine if choice of extinction law affects our spectral fitting results. However, these various extinction laws did not change much our recovered spectral types or best-fit $\chi^2$ values. The alpha=-2.1 law produced a $<$ 0.5 sub-type shift to later spectral type for  all targets.

Our initial confirmation rate for $Q$-index selected objects in Serpens-South stands at 4 out of 7 candidates, however, all three early M contaminants lie extremely close to the edge of   one of the four WIRCAM array detectors; upon re-examination of the filter images, it became clear that the Q values calculated for these objects are spuriously low due to cosmetic issues  at the edge of the detector.  For further spectroscopic followup (as described in Dubber et al. in prep), we have avoided following up similar spurious detections at the edge of the detector.  \\

\begin{figure*}
\centering
\includegraphics[scale=0.8, angle = 0]{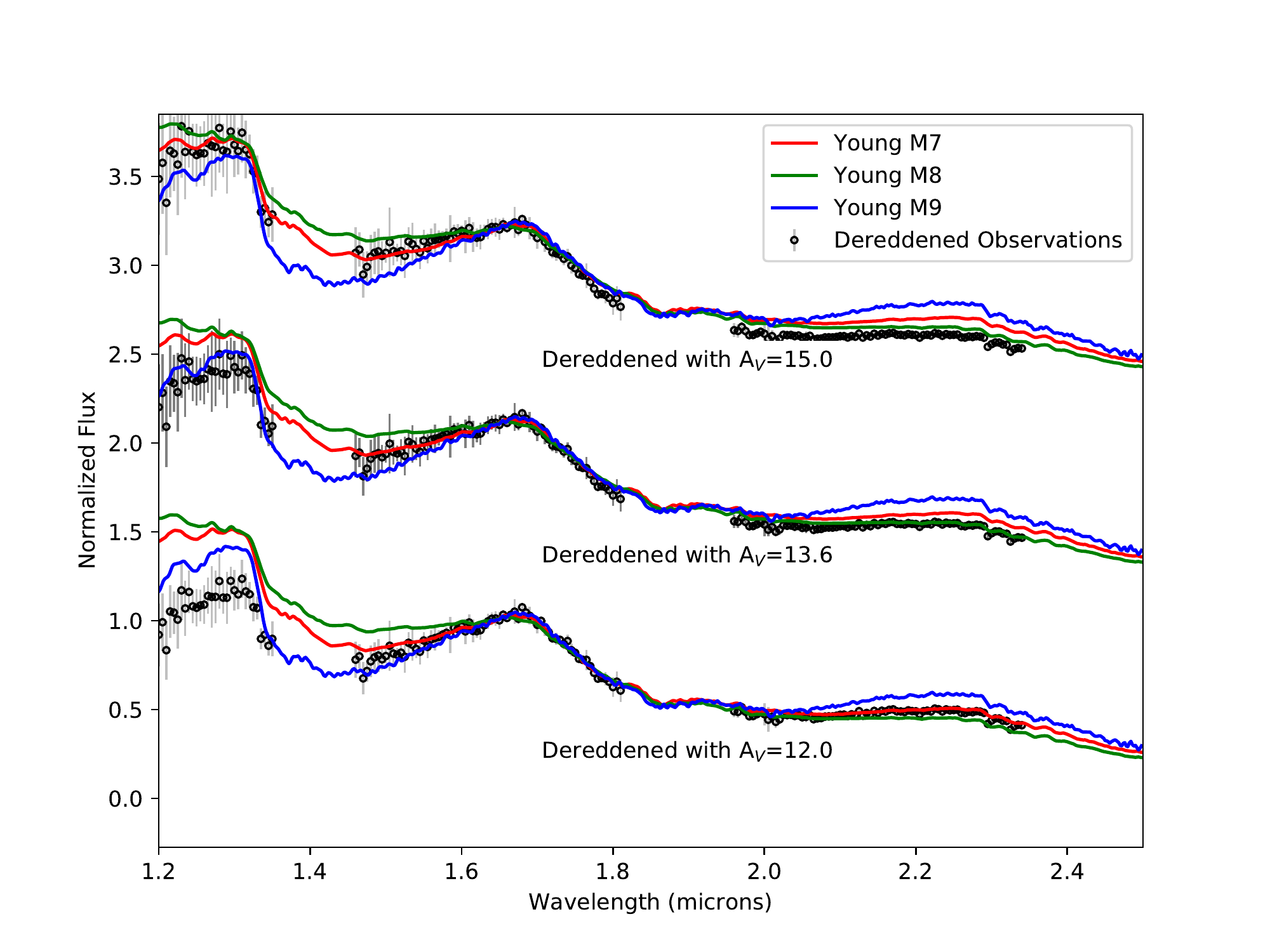}
\caption{ Normalized, dereddened spectra of SERP183044-020918 along with  three best fit  spectral templates and the respective 
$A_V$ values. Spectra has been binned onto a regular grid sampled at every 5 nm, down from roughly 0.2 nm at full resolving power. The grey error bars show the 1-sigma uncertainty of the binned data.}
\label{spectype_4}
\end{figure*}

\begin{figure*}
\centering
\includegraphics[scale=0.8, angle = 0]{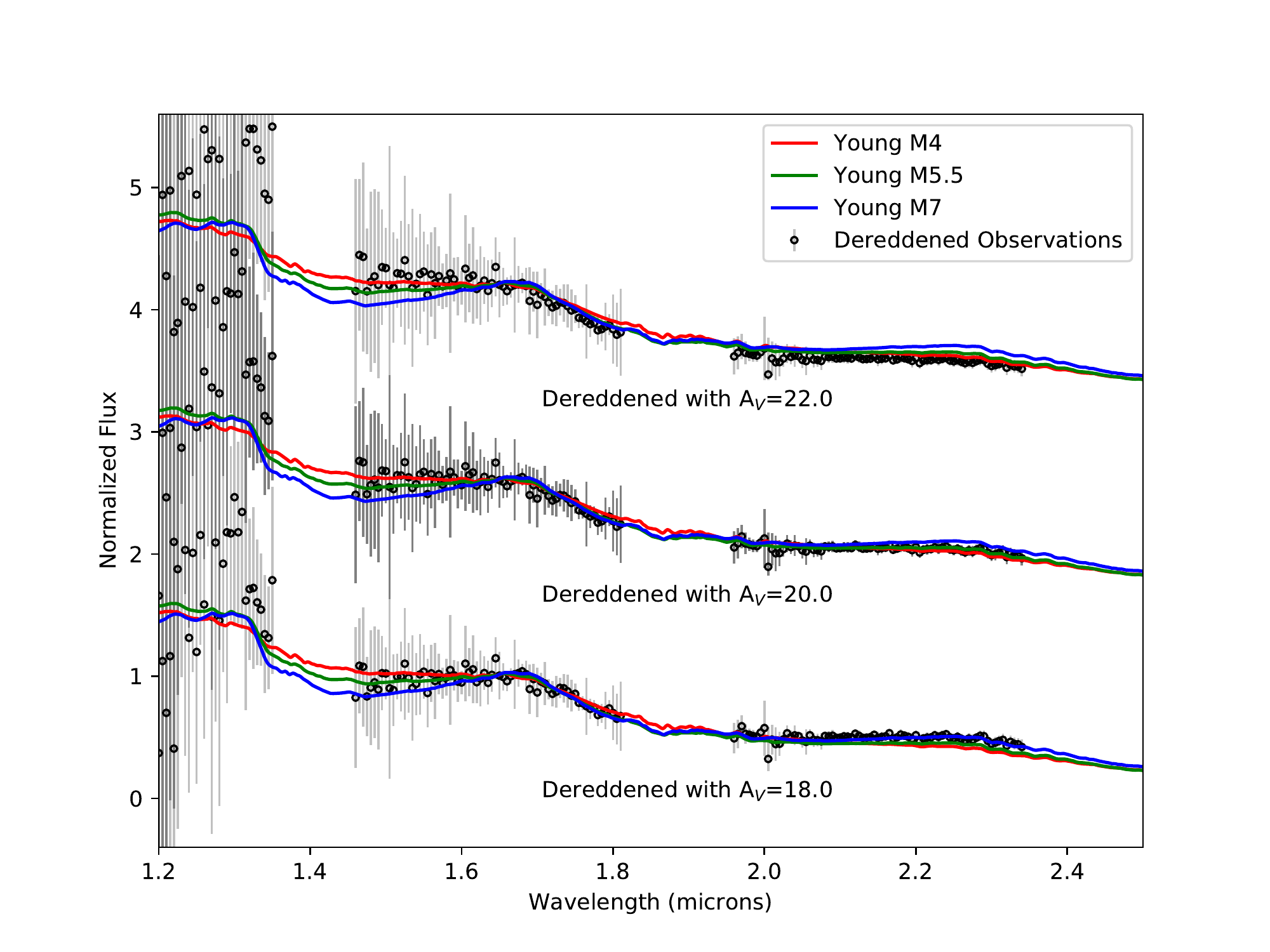}
\caption{ Normalized, dereddened spectra of   SERP182917-020340 along with  three best fit  spectral templates and the respective 
$A_V$ values.  Spectra has been binned onto a regular grid sampled at every 5 nm, down from roughly 0.2 nm at full resolving power}. The grey error bars show the 1-sigma uncertainty of the binned data.
\label{spec_5}
\end{figure*}

\begin{figure*}
\centering
\includegraphics[scale=0.8, angle = 0]{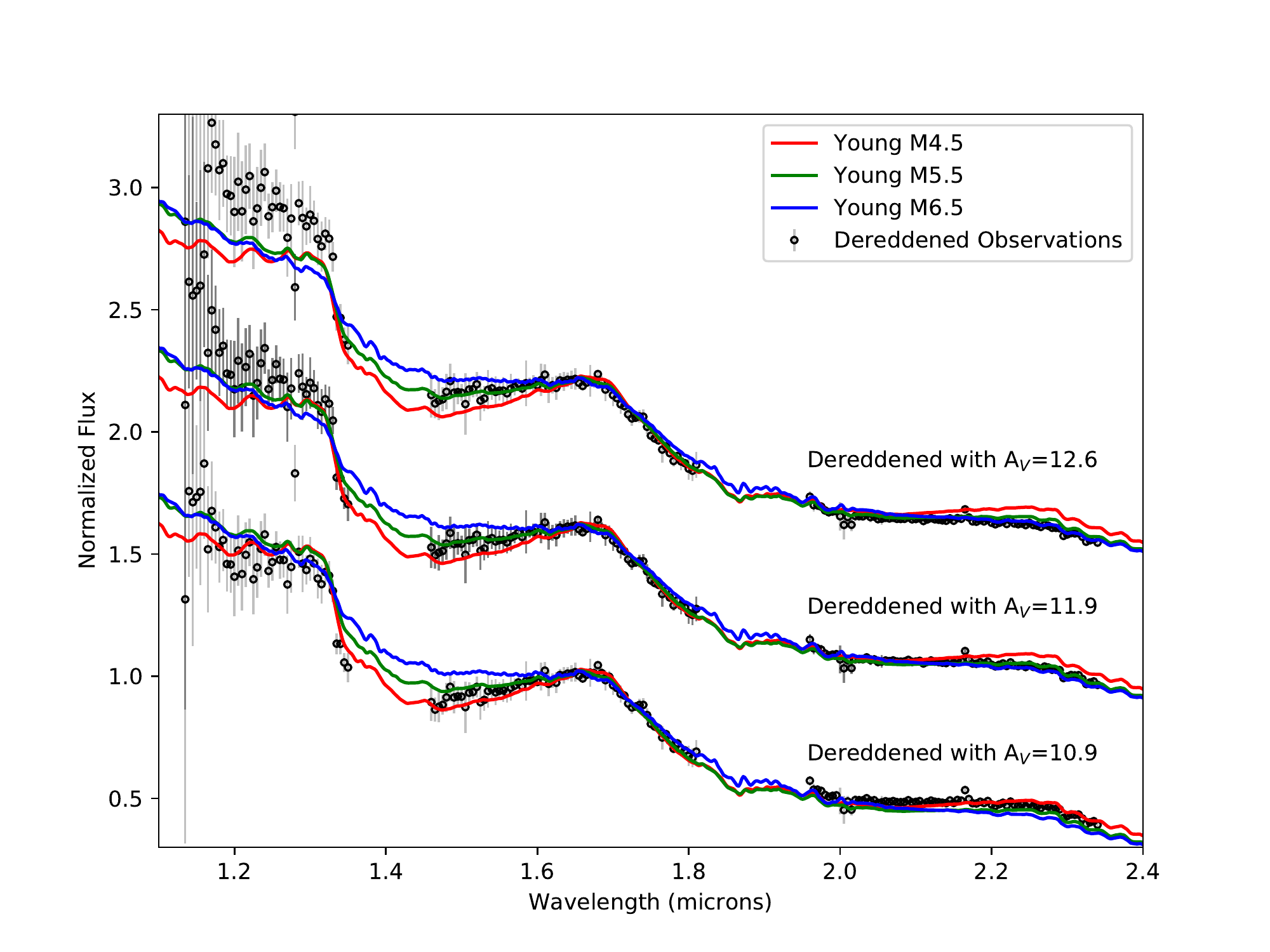}
\caption{ Normalized, dereddened spectra of SERP182918-020245 along with  three best fit  spectral templates and the respective 
$A_V$ values.  Spectra has been binned onto a regular grid sampled at every 5 nm, down from roughly 0.2 nm at full resolving power}. The grey error bars show the 1-sigma uncertainty of the binned data.
\label{spec_6}
\end{figure*}

\subsection{New candidate low-mass, young members of Serpens-South}

The spectra of young  objects later than M5 show a characteristic triangular peak in the H-band \citep{cushing2005}. Of the seven dereddened  spectra of candidate low-mass objects, the H-band pseudo-continuum  of SERP183044-020918, SERP182917-020340 and SERP182918-020245 have a definitely peaked shape, and SERP183038-021419 shows hints of a possibly peaky H-band (see Figs. \ref{spec_2} to \ref{spec_6}). This feature, where the H-band is dominated by 
strong water absorption bands to either side of the sharp peak located between 1.68 and 1.70 $\mu$m, was first reported by \citet{lucas2001}
for very young ($\sim$ 1 Myr), cool objects in the Trapezium in Orion. The triangular shaped peak is  ascribed to the effects of 
low pressure and low gravity atmospheres typical of self-gravitational collapsing objects with $T_{eff}$ $\le$ 2700 K  and is a  
consistent signature of youth (e.g., \citealt{allers2007,rice2010,allers2013,liu2013, schneider2014,gagne2014,esplin2017}). The 
triangular shaped H-band continuum of these four objects confirms that they are young and likely members of the Serpens-South star 
forming region.

\subsection{New members with circumstellar disks}
\label{excess}
Serpens-South is known to have a high fraction of young stellar objects hosting circumstellar disks 
(\citealt{gutermuth2008,dunham2015}). The mid-IR photometry from {\it Spitzer} and {\it WISE} can be used 
to identify the presence of circumstellar disks around low-mass young objects (\citealt{luhman2008a,luhman2008b}). 

\begin{figure}
\centering
\includegraphics[scale = 0.9, trim = 0 0 20 0, clip]{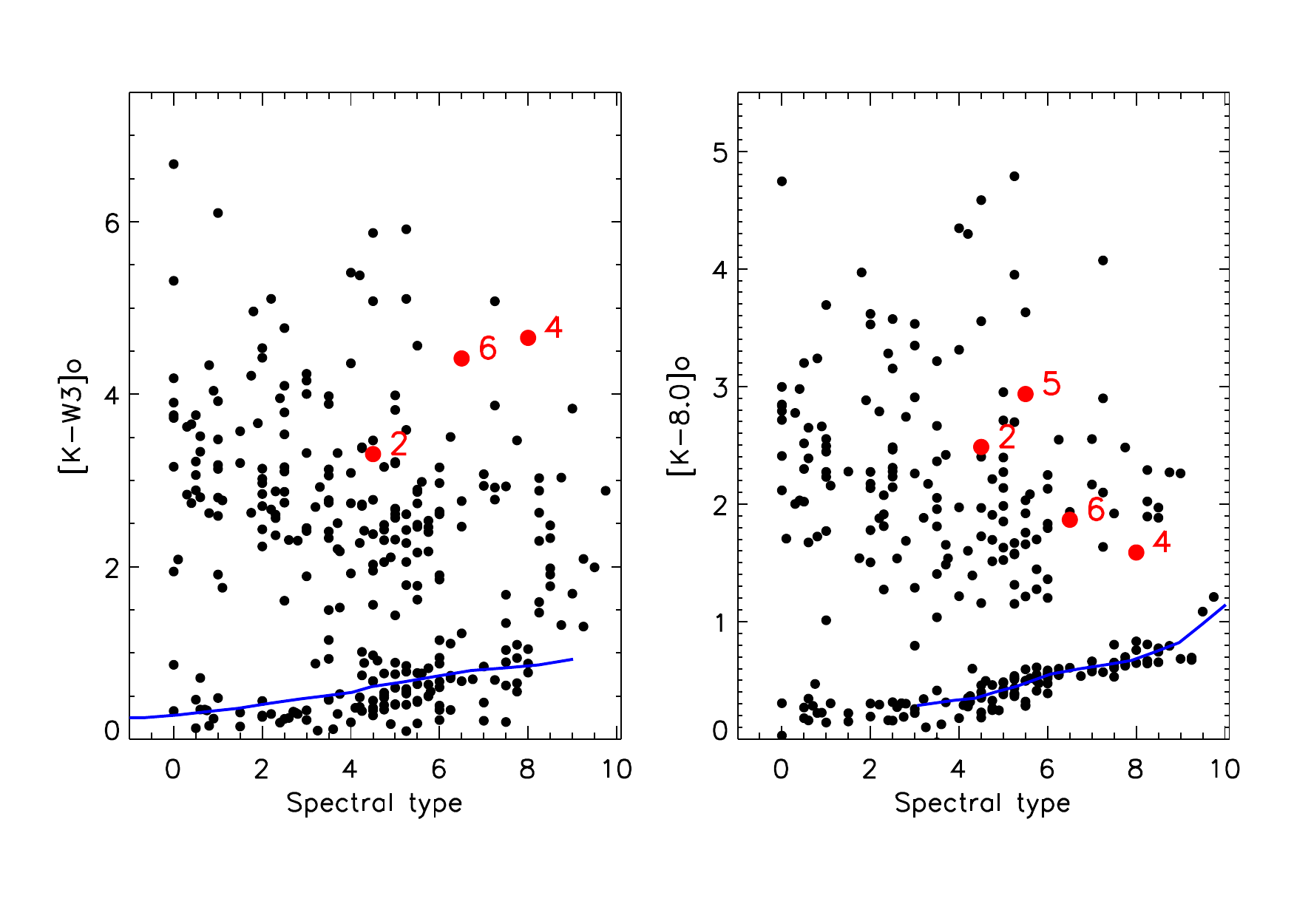} 
\caption{Extinction corrected mid-IR colors as a function of spectral type (M0 (0) to  L0 (10))  for late type members 
of Taurus from \citet{luhman2017} and \citet{esplin2014} (black circles)  and the newly identified low-mass members of 
Serpens-South are shown in red. The intrinsic photospheric colors for young objects are represented by the blue curve \citep{esplin2017}.  
} 
\label{colorexcess}
\end{figure}

To determine whether the four  candidate low-mass members of  Serpens-South  have color excesses from disks in 
the {\it Spitzer} and {\it WISE} photometry, we compared their dereddened $[K-W3]$, $[K-W4]$, $[K-5.8]$ and $[K-8.0]$  colors   with the typical 
photospheric colors of young stellar objects given in \citet{luhman2010} and  \citet{esplin2017}.  The reddening of 
these targets are obtained from the spectral fitting procedure described in section \ref{reddening}. Though the presence of 
disks can artificially redden the stars, we ignore this factor as it is beyond the scope of this paper. 
 In figure \ref{colorexcess} we compare the $[K-W3]$ and $[K-8.0]$ colors of the low-mass members of Taurus \citep{luhman2017,esplin2014} 
with the new members identified in Serpens-South. 
The dereddened  colors of the  low-mass candidates have a  $>$ 1-2 mag excess above their typical photospheric 
colors given in \citet{esplin2017}. Similarly, the dereddened colors in [3.6 - 5.8] and [3.6 - 8.0] are 
significantly redder than their corresponding photospheric sequence \citep{esplin2017} for the above four low-mass 
targets. The sizes of the excesses of these targets are similar to those of members of Cha I (\citealt{luhman2008a,luhman2008b}) 
and Taurus \citep{esplin2014} with disks, and hence the  reddened  colors of four late type  Serpens-South members indicate 
that they are likely to host disks. 

The presence of circumstellar disks provides additional evidence of the youth and membership of those objects.  
Additionally, the above four targets, ie., SERP183044-020918, SERP182918-020245, SERP183038-021419 and SERP182917-020340 
are listed as young stellar objects in the  SFiNCs Xray-Infrared catalog by \citet{getman2017} as well as in 
Gould Belt Survey by  \citet{dunham2015}. The  IR excess confirm the youth and likely membership of 
these candidate low mass targets in Serpens-South.

\subsection{Ongoing accretion in SERP182918-020245}
\label{accretion}

\begin{figure}
\centering
\includegraphics[scale = 0.8, trim = 0 0 0 0, clip]{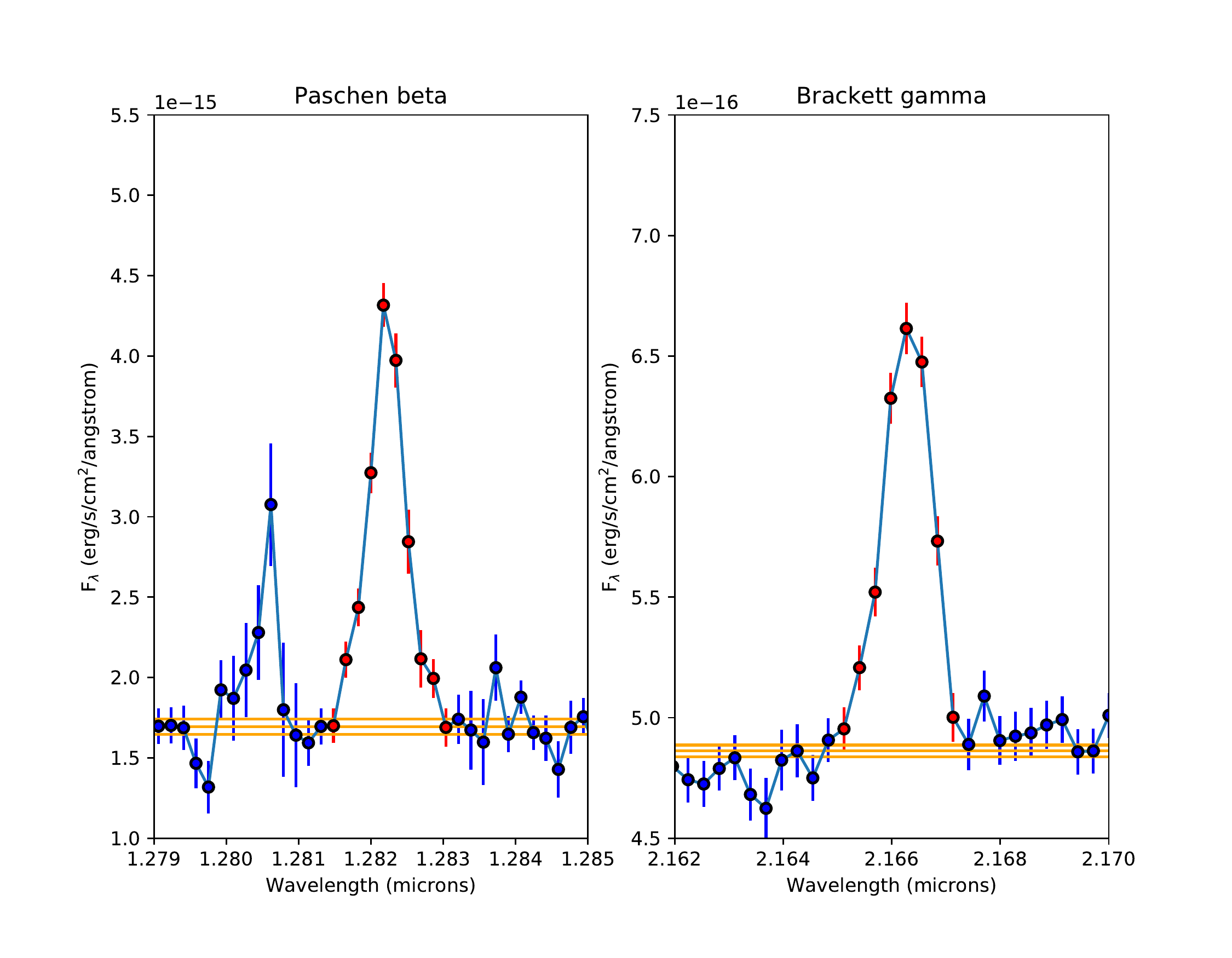} 
\caption{$Pa{\beta}$ and  $Br{\gamma}$ emission lines identified on the dereddened spectra of SERP182918-020245. Line fluxes were measured by summing the red data points after subtraction of the continuum level  (middle orange lines with its $\pm1\sigma$ uncertainty).}
\label{serp_2lines}
\end{figure}

Magnetospheric accretion in young stellar objects produces emission lines that span from UV to IR wavelengths.  
These emission lines can be used to obtain an estimate of the accretion luminosity, and are also signatures of the chromospheric activity of the star. In the near-IR regime, $Pa{\beta}$ at $\lambda$=1.2818 $\mu$m and $Br{\gamma}$ at 
2.16 $\mu$m  are the major  emission lines present. One of the four candidate low-mass members in our list,  SERP182918-020245,
exhibits $Pa{\beta}$ and $Br{\gamma}$  emissions  (see Fig. \ref{serp_2lines})   which confirms its youth and association 
with Serpens-South. The fluxes  of the $Pa{\beta}$ and  $Br{\gamma}$  emission lines are   computed by directly 
integrating the flux within the Gaussian curve along these lines. We used the flux calibrated,  extinction-corrected 
spectra for this analysis. The line luminosities of these two emission lines are computed using the relation
$L_{line}$ =  4$\pi$ $d^2$ $f_{line}$, \citep{alcala2014}, where, $d$ is the distance to the target, i.e., 436$\pm$9.2 pc and $f_{line}$
is the extinction corrected flux. 
Taking into account the range of $A_V$ values producing acceptable spectral fits and again adopting the \citet{cardelli1989} reddening law, we find extinction corrected fluxes of  1.64$^{+0.52}_{-0.62}$ $\times$ 10$^{-14}$  ergs s$^{-1}$ cm$^{-2}$ and 2.00$^{+0.24}_{-0.38}$ $\times$ 10$^{-15}$  ergs s$^{-1}$ cm$^{-2}$ respectively for the $Pa{\beta}$ and  $Br{\gamma}$ emission lines.
 We calculate the accretion luminosity of SERP182918-020245 using the $L_{acc}$ versus 
$L_{line}$ relationships for the $Pa{\beta}$ and  $Br{\gamma}$ emission lines given in \citet{alcala2014}.  The  
estimated $log (L_{acc} / L_{\odot})$ values from these two lines are -1.72$\pm$0.53 and -2.11$\pm$0.52 respectively.

\subsection{Physical parameters of the low-mass objects}
\begin{figure}
\centering
\includegraphics[scale = 0.9, trim = 0 0 200 0, clip]{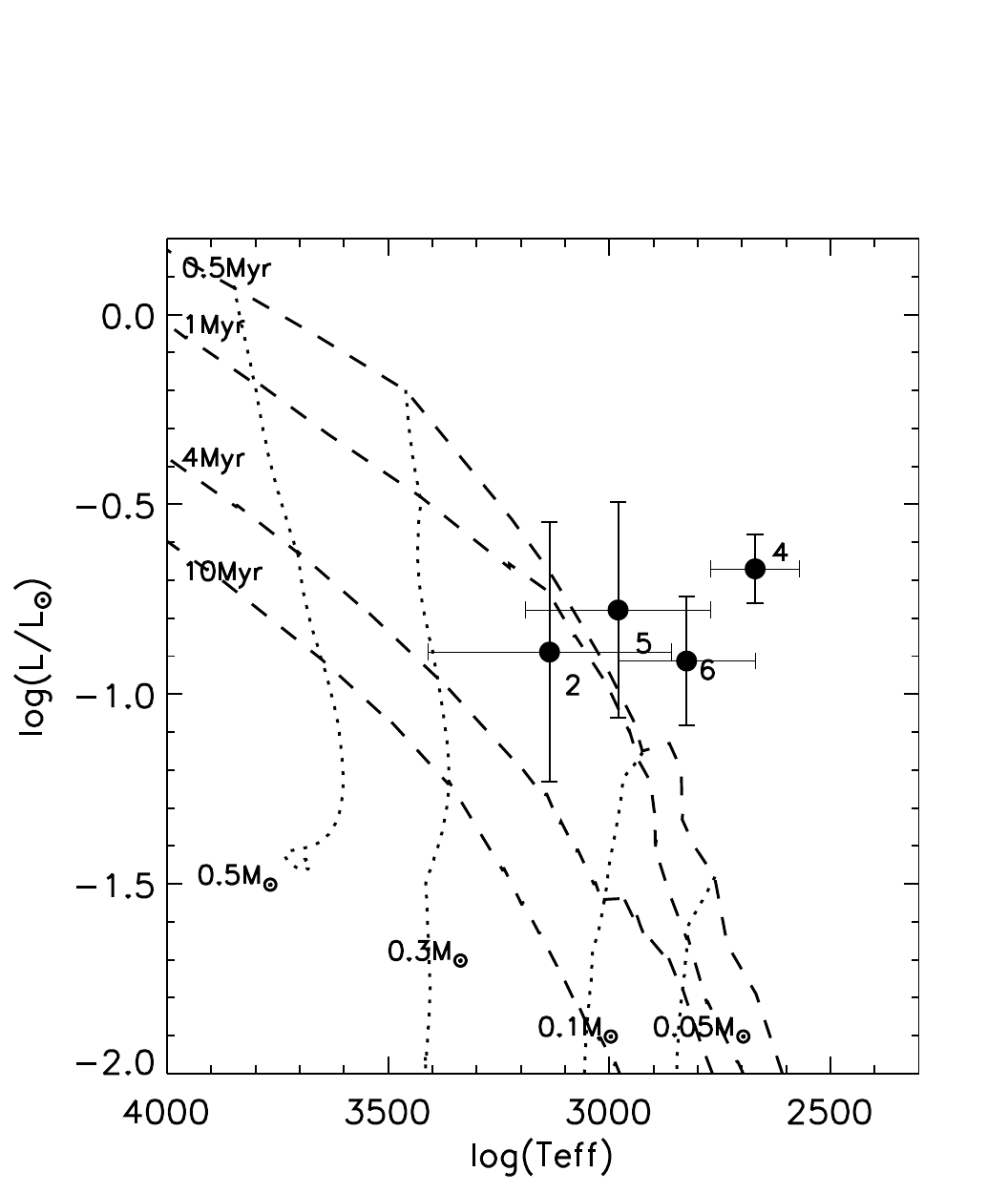} 
\caption{H-R diagram for the four low-mass candidate members identified in Serpens-South. Isochrones (dashed curves) for 0.5, 1, 4 
and 10 Myr ages and   evolutionary tracks (dotted curves) for masses 0.5, 0.3, 0.1 and 0.05 M$_{\odot}$ taken from \citet{baraffe2015} 
are also shown. }
\label{hrd}
\end{figure}

In order to estimate the physical parameters of the candidate low-mass members  identified in Serpens-South,  we used the apparent magnitudes listed in Table~\ref{table:phot}. For bolometric corrections in  $J$ and $K$- bands, we used the values derived by 
\citet{herczeg2015} and \citet{filippazzo2015} for young M type objects. We adopted the latest estimate on  distance to the 
Serpens star forming region i.e., 436 $\pm$ 9.2 pc (see section \ref{intro}), which corresponds to a distance modulus of 
8.2 $\pm$ 0.04 mag.  Using the extinction values in $V$-band obtained  from the best fitting template spectral types  
(see section \ref{reddening}) and   adopting the reddening law from \citet{cardelli1989} (ie., $A_J$ / $A_V$ = 0.282 
and $A_K$ / $A_V$ = 0.112), the apparent magnitudes are corrected for reddening.   
For the Sun, we adopted $M_{bol}$ = 4.73 mag. The luminosities thus calculated from $J$ or $K$-band magnitudes for the four 
candidate low-mass objects are given in Table \ref{phypars}. The uncertainty in luminosity is estimated from the quadratic 
sum of  fractional errors in distance, photometry, extinction and bolometric correction. The final uncertainty in each target is dominated 
by the error associated with the  extinction measurement, which is relatively high towards Serpens-South (See Table \ref{phypars}). 
The corresponding effective temperature for each spectral range is obtained from \citet{herczeg2014}. Assuming an age of $\sim$ 0.5 Myr for Serpens-South \citep{gutermuth2008}, the luminosities are then compared to the 0.5 Myr isochrone from \citet{baraffe2015} to estimate the masses  
of the targets. Fig. \ref{hrd} shows the HR-diagram of the four low-mass members along with the isochrones and evolutionary 
tracks from \citet{baraffe2015}. 

Two of  the four low-mass targets (specifically ID numbers 4 and 6) shown in Fig.~\ref{hrd} appear significantly brighter than expected from evolutionary model predictions for 0.5 Myr objects. 
Similar trends have been observed in other  star forming regions such as Taurus, Ophiuchus, Chamaeleon etc.  (eg. \citealt{zhang2018, ribas2017,muzic2013,herczeg2015}). However, the masses and luminosities inferred for $\leq$1 Myr objects are very uncertain and strongly model dependent.  
The initial radius of a young object at its birth location depends on the initial entropy and the entropy gained and lost during the main accretion phase \citep{stahler1983,hartmann1997,hosokawa2011}.  Thus, the early accretion history of young low-mass stars and brown dwarfs will affect the physical properties of objects (luminosity, radius, effective temperature etc.) in star forming regions and young clusters (although the effect of this accretion history will become negligible by ages $>$10 Myr).  This accretion history, however, is generally not well-known for any given object, and can vary significantly from object to object in a given region.   Modelling by \citet{baraffe2017} demonstrates that accretion history produces a significant spread in luminosity on HR  diagrams for a coeval population of young low-mass stars and brown dwarfs.  The location of our new candidate Serpens members above the 0.5 Myr isochrone suggests that they were born with higher entropy (and thus a larger initial radius) than predicted from standard models of evolution, which do not take into account the individualized accretion history of any given object \citep{baraffe2015}.  Binarity could also be a reason for their higher luminosity beyond the model isochrones. Nonetheless, targets 2 and 5 have effective temperatures consistent with low-mass stars of mass $\sim$ 0.1 $M_\odot$ whereas targets 4 and 6 have effective temperatures consistent with substellar objects with mass $<$ 0.08-0.05 $M_\odot$.  These are the coolest and lowest mass members of Serpens-South star forming complex identified so far. 


\begin{table*}
\centering
\caption{Physical parameters of  the candidate low-mass Serpens-South members}
\label{phypars}
\begin{tabular}{ccccccccc}
 \hline
ID & Name& Spectral type &  Av  & $BC_J\footnote{Bolometric correction from \citet{herczeg2015}}$ &  $BC_K\footnote{Bolometric correction from \citet{filippazzo2015}}$  & log $L/L_{\odot}$ & $ T_{eff}$   \\
&    & & mag  & mag & mag              &  &   (K)     \\
\hline
2 &  SERP183038-021419 & M3-M6          & 18-24 &1.84-2.03& &-0.89$\pm$0.34 &3135$\pm$275  \\
4 &  SERP183044-020918 & M7-M9          & 12-15 & & 3.07$\pm$0.13&-0.67$\pm$ 0.09 &2670$\pm$100  \\
5 &  SERP182917-020340 & M4-M7          & 17-22 &1.93-2.06 & &-0.78$\pm$0.28 &2980$\pm$210  \\
6 &  SERP182918-020245 & M5-M8          & 10-13 &1.99-2.06 & &-0.91$\pm$0.17 &2825$\pm$155  \\

\hline
\end{tabular}
\end{table*}


\section{Discussion}

Serpens-South is a highly extincted and extremely dense region, lying right along the Galactic plane.  It is an extremely challenging site to study due to the high number of reddened background contaminants with which we must contend.  The W-band technique has enabled us to overcome these limitations and discover the first likely substellar members of this cluster, but the extreme extinction along this line of sight pushes the W-band technique to its limits.  As noted previously, the major uncertainty driving the error in our spectral types is from the fit to A$_V$ -- quite a wide range of A$_V$ values fit well.  The $Q$ index we have defined is also set to $e$ values appropriate for moderately extincted regions -- specifically, we set the value of $e$ as appropriate for an M0 reddened with an $A_V$ of 10.  In comparison, most of our candidate members have $A_V$ $>$ 15.  

The Q-index that we have constructed is solely a measure of spectral type (based on the depth of water absorption features), not age -- it will just as easily identify background or foreground late-M field stars as well.  This renders spectroscopic followup imperative.  Three out of four of our candidate new Serpens-South members show "peaky" H band spectra, a clear indicator of the youth of these objects.  All 4 candidate members have additional corroborating evidence of youth from color excesses indicating circumstellar disks.  One additionally has clear accretion features in its spectrum.  Thus, we are certain that our 4 candidate members are very young ($<$2 Myr).  Nonetheless, we cannot fully confirm them as members of Serpens-South.  Full confirmation would require demonstrating that these candidates share common proper motions and radial velocities with Serpens-South stars \citep[e.g.~as~done~for~closer~and~somewhat~older~moving~group~stars~in~][]{gagne2014}.  Unfortunately, at a distance of $>$400 parsec and with extreme extinction ranges  the objects are too faint at optical wavelengths for Gaia astrometry.   

\section{Conclusions}

Serpens-South is an active star forming complex in the solar neighborhood, yet the low-mass members of the region remain largely undiscovered, due to the high extinction towards this region. In this paper we present the preliminary results of our survey using  the novel W-band filter at the WIRCam/CFHT and follow-up spectroscopic observations to identify and characterize the sub-stellar members of Serpens-South.  Further spectroscopic followup of fainter candidates and a calculation of the IMF for Serpens-South will be presented in Dubber et al. in prep.  After screening the probable low-mass candidates based on their W-band based Q-index, we obtained spectra for the 7 brightest targets among them. Of these, four are probable members of Serpens-South.  Three of these objects have spectral types later than M4 spectral type and evidence of youth from peaky H-band pseudo-continuum spectral morphology.  All four objects have IR excess emission, indicating the likely presence of circumstellar disks around them \citep{getman2017, dunham2015}. One of the four candidate low-mass members in our list, SERP182918-020245, exhibits  $Pa{\beta}$ and $Br{\gamma}$  emission  which confirms its youth and ongoing magnetospheric accretion.  These are the coolest and lowest mass candidate members yet identified in Serpens-South.
\\

\acknowledgments

Based on observations obtained with WIRCam, a joint project of CFHT, Taiwan, Korea, Canada, France, at the Canada-France-Hawaii Telescope (CFHT) which is operated by the National Research Council (NRC) of Canada, the Institut National des Sciences de l'Univers of the Centre National de la Recherche Scientifique of France, and the University of Hawaii.  Visiting Astronomer at the Infrared Telescope Facility, which is operated by the University of Hawaii under contract NNH14CK55B with the National Aeronautics and Space Administration.  The authors wish to recognize and acknowledge the very significant cultural role and reverence that the summit of Maunakea has always had within the indigenous Hawaiian community.  We are most fortunate to have the opportunity to conduct observations from this mountain. JJ acknowledges the  grant no. 11473005 awarded 
by the National Science Foundation of China to GJH with which most of work was carried out and also to Telescope Access Program (TAP) 
through which CFHT time was awarded to pursue part of the W-band observations. 


\bibliographystyle{apj}
\bibliography{ref_wband}

\clearpage
\end{document}